\documentstyle[12pt]{article}
\font \tbfontt                = cmbx12
\font \tbfonts                = cmbx7  at 8.4 pt
\font \tbfontss               = cmbx5  at 6 pt
\newfam\cmbxfam
\textfont\cmbxfam=\tbfontt \scriptfont\cmbxfam=\tbfonts
\scriptscriptfont\cmbxfam=\tbfontss
\def\bf{\fam\cmbxfam\tbfontt}
\begin{document}
\parskip = 0 pt
\baselineskip = 16 pt
\parindent = 20 pt
\centerline{\LARGE  Time scale, objectivity and irreversibility}\par
\centerline{\LARGE  in quantum mechanics}\par
\vskip 15  pt
\centerline{L.~Lanz\footnote{Dipartimento di Fisica
dell'Universit\`a di Milano and Istituto Nazionale di Fisica
Nucleare, Sezione di Milano, Via Celoria 16, I-20133, Milan,
Italy. E-mail: lanz@mi.infn.it} and O.~Melsheimer\footnote{Fachbereich
 Physik Philipps-Universit\"at,
 Renthof 7,D-35032 Marburg,
Germany. E-mail: melsheim@mailer.uni-marburg.de}}
\vskip 15 pt
\par
\section{Introduction}
\par
For the physical interpretation of the formalism of quantum mechanics (QM)
a classical world of sources and detectors of microsystems must be invoked in
order to get rid of the so-called E.R.P. paradox and the subjectivity
 within the measuring process.
Substituting invocation with high mathematical effort, Ludwig obtained
 QM as the description of a microphysical interaction channel between
 macrosystems and showed that QM must be enriched by the mathematical tools
 that are now called POV measures, operations and instruments  (for a
more recent overview see (Ludwig, 1983)).
It is very remarkable that these concepts also arose in more accurate
descriptions of the measuring process and more profound thinking about the
statistical  structure of QM (Davies, 1976), (Holevo, 1982) and (Kraus, 
1983).
However the problem remains to reconcile the classical phenomenological
description of systems with the quantum mechanical description of their
microphysical structure. Our aim is to reconsider this problem taking
in an explicit way isolated systems as basic elements of reality and looking
at QM, already in its field formulation, 
as the basic theory of finite isolated systems;
only afterwards one arrives at particles when the peculiarities of finite
systems, like boundary conditions, are negligible: in this philosophy
the thermodynamic limit has just a reversed role, it is not used in order
to reveal the classical behavior of macrosystems, but is necessary to
attain local covariance and universality of the theory. All the difficulties
related to quantum mechanical inseparability, enter now in the very concept
of isolated system: our point of view implies a weakening of the idea
of an absolutely isolated system. Isolation is relative to a suitable
set of {\sl slow variables}, whose dynamics restricted to expectations, as
phenomenology indicates, has only a restricted memory of the previous values of
these expectations.
\par
So to achieve isolation one has to establish a suitable time scale, choose
variables with expectations having a typical variation time of this scale,
prepare the system inside some confined space region during a suitable
time interval, controlling and measuring the relevant variables inside
a suitable preparation time interval; we have to explicitly introduce
in the formalism the fact that restricting to suitable variables and using
some effective description of quasilocal field interaction, the too remote
history of the system has to be neglected. On the contrary if one pretends
to describe the local behavior with completely sharp time specification
one can  expect that the full history of the whole universe would be 
involved.
\par
By taking finite isolated systems, instead of particles as the main subjects,
the typical
ultraviolet and infrared problem of quantum field theory is absorbed inside the
non universal features of the description. The opening to
irreversibility entailed by this point of view, will be discussed in Section 3 
to settle problems arising in the classical description of a 
macrosystem; such a description is given in Section 2, essentially by a suitable
reconsideration of Zubarev's  approach to non equilibrium 
thermodynamics. As an example Boltzmann description of a dilute gas is 
discussed in Section 4.
\par
\section{Description of a finite isolated system}
\par
The microphysical structure  is described by a set of interacting
quantum Schr\"odinger fields. Here we consider the simplest model: one
interacting quantum Schr\"odinger field  (QSF) 
$\hat\psi_{nc}({\mbox{\bf x}})$ (not yet
confined) , to which  the following local Hamiltonian density is associated:
\begin{eqnarray}
\label{1.1}
\lefteqn{
{\hat e}_{nc}({\mbox{\bf x}})
  =
 {{
 \hbar^2
   \over
   2m
   }}
   \nabla
   {\hat \psi}^{\scriptscriptstyle\dagger}_{nc}({\mbox{\bf x}})
   \cdot
   \nabla
   {\hat \psi}_{nc}({\mbox{\bf x}})+ {} }
   \\
   & {} +
   {1\over 2}
   \int d^3\! {\mbox{\bf  r}}
   \,
   {\hat \psi}^{\scriptscriptstyle\dagger}_{nc}
   ({\mbox{\bf x-\bf r/2}})
   {\hat \psi}^{\scriptscriptstyle\dagger}_{nc}
   ({\mbox{\bf x+\bf r/2}})
   V(|{\bf r}|)
   {\hat \psi}_{nc}
   ({\mbox{\bf x+\bf r/2}})
   {\hat \psi}_{nc}
   ({\mbox{\bf x-\bf r/2}})\nonumber
   \end{eqnarray}
   \begin{equation}
   \label{1.2}
   {
   \left[
   {\hat \psi}({\mbox{\bf x}}),
   {\hat \psi}^{\scriptscriptstyle\dagger}({\mbox{\bf x}}')
   \right]
   }_\pm
     =
   \delta ({\mbox{\bf x}}-{\mbox{\bf x}}').
   \end{equation}
The basic local dynamical variable of this model

	\[
	 {\hat\psi}_{nc}({\bf x},t)=e^{\frac{i}{\hbar}
	  H_{nc} t}{\hat\psi}_{nc}({\bf x})e^{-\frac{i}{\hbar} H_{nc} t }
	 \quad , \quad 
          \hat H =\int_{\omega} d^{3} {\bf x}  \hat e_{nc}({\bf x})
          \]
satisfies the Schr\"odinger field equation

	 \begin{eqnarray}
	 \label{1.3}
	 \lefteqn{
	 i{\hbar}\frac{\partial{\hat\psi}_{nc}({\bf x},t)}{\partial t}
	  =-\frac{{\hbar}^{2}}{2m}
	 \Delta{\hat\psi}_{nc}({\bf x},t)+ {}}
	 \\
	 & & {} +    \int d^{3}{\bf y} {\hat\psi}_{nc}^\dag({\bf y},t)
	  V(|{\bf x}-{\bf y}|){\hat\psi}_{nc}({\bf y},t){\hat\psi}_{nc}
	   ({\bf x},t) \nonumber
	  \end{eqnarray}
 As it is well known and will appear in the results, choice 
(\ref{1.1}) means
  in usual language,
N-body system of structureless molecules interacting via the two body potential
 $V|{\bf x}_{1}-{\bf x}_{2}|)$.\\
 The field equation (\ref{1.3}) that accounts for covariance under Galilei
  transformations, has for the massive
 continuum we are describing a similar role that Maxwell equations have for
 electromagnetism, however, due to the irrelevance of 
self-interaction in the electromagnetic
 case, the classical field theory plays in the latter case a much more extended
  role.
  \par
 Confinement of the system inside a region $\omega\subset {\mathbf R}^{3}$ is
 obtained expanding ${\hat\psi}({\bf x})$ on the normal modes $u_{f}({\bf x})$
 of the system, where
 \begin{displaymath}
 -\frac{{\hbar}^{2}}{2m} \Delta u_{f}({\bf x})=E_{f}u_{f}({\bf x})
 \qquad   u_{f}({\bf x})=0 \qquad {\bf x} \in
 \partial\omega
 \end{displaymath}
 setting
 \begin{equation}
 \label{1.4}
 {\hat\psi}({\bf x}) =
	\sum_f u_f ({\mbox{\bf x}}) {\hat
	a}_f   \qquad   \left [{\hat a}_{f},
{\hat a}_{g}^\dag \right ]_\pm=\delta_{fg}
	\end{equation}
We replace ${\hat\psi}_{nc}({\mbox{\bf x}})$ by   $\hat\psi({\bf x})$ for
 ${\bf x}
\in\omega$  and by $0$ for ${\bf x}\not\in \omega$,
 in the energy density $\hat e({\bf x})$ and in all
relevant expressions built with the field operators. In this way peculiar 
confinement
is superposed to the quasilocal universal (within the limits related
to the effective potential $V(|{\bf r}|)$) behavior.\nonumber \\
Our aim is not at all a full description of the finite isolated system,
but to give a description of it having negligible correlations with the
environment; this description is related to suitable {\sl {slow 
variables}}, 
linked to the fundamental
constants of motion of the system: mass and energy.\nonumber \\
The densities generating these observables are the {\sl {relevant 
variables}}; as 
phenomenology
indicates there are two meaningful descriptions:\nonumber  \\
A:  the hydrodynamic one based on energy density and mass density,\nonumber \\
B: the kinetic one based on energy density and phase-space density;
 \nonumber \\
  the energy
density is given by equation (\ref{1.1}) 
with ${\hat\psi}_{nc}({\bf x})$ replaced
 by $\hat\psi({\bf x})$, the  mass
density is given by
\begin{equation}
\label {1.5}
\hat m({\bf x})=m{\hat\psi}^\dag({\bf x})\hat\psi({\bf x})
\end{equation}
the phase-space density is given by:
\begin{equation}
\label{1.6}
\hat f({\bf x},{\bf p})=\sum_{hk}{\hat a}_{h}^\dag<u_{h}|\hat F^{(1)}({\bf x},
{\bf p})|u_{k}>\hat a_{k}
\end{equation}
$\hat F^{(1)}({\bf x},{\bf p})$ being the density of a POV measure
 for the joint position-momentum measurement in one-particle quantum
 mechanics (Lanz, Melsheimer and Wacker, 1985).
\par
The first step towards a classical description is the axiomatic introduction 
of a velocity field of the continuum,
so that the former observables can be referred to a local rest frame:
denoting by an index $^{(0)}$ these observables, one has:
	\begin{eqnarray}
	{\hat e}^{(0)}({\bf x})=
	 \frac {1}{2m}
	\left(  
	i\hbar \nabla - m{\bf v}
	({\bf x},t)
	\right)  
	{\hat \psi}^{\dag}({\bf x})
	\cdot  
	\left(  
	-i\hbar \nabla -
	m{\bf v}({\bf x},t)
	\right)  
	{\hat \psi}({\bf x}) + {}
		      \nonumber \\
       {} + \frac 12
	\int d^3\! {\bf  r}
	\,  
	{\hat \psi}^{\dag}
	({\bf x}-\frac{{\bf r}}{ 2})
		{\hat \psi}^{\dag}
	({\bf x}+\frac{{\bf r}}{ 2})
	V(|{\bf r}|)
	{\hat \psi}  
	({\bf x}+\frac{{\bf r}}{ 2})
	{\hat \psi}  
	({\bf x}-\frac{{\bf r}}{ 2})
	\nonumber
	\end{eqnarray}
	\begin{equation}
        \label{1.7}
	 {\hat m}^{(0)}({\bf x})=\hat m({\bf x})
	\end{equation}
	\[
	 \hat f^{(0)}({\bf x},{\bf p})=
\hat f({\bf x},{\bf p}-m{\bf v}({\bf x},t))
	 \]
  The introduction of this external classical field allows to compensate
    a gauge transformation of the field 
 $ \hat\psi({\bf x}) \rightarrow \hat\psi({\bf x})
  e^{\frac{i}{\hbar}\Lambda({\bf x})}$
  with a transformation 
${\bf v}({\bf x},t)\rightarrow{\bf v}({\bf x},t) - 1/m 
\nabla \Lambda({\bf x},t)$
  of the external parameter. ${\bf v}({\bf x},t)$ is linked to the expectation
  at time $t$ of the momentum density of the system:
	\begin{eqnarray}
	\label{1.8}
	{\hat {\mbox{\bf p}}}({\mbox{\bf x}})
	&=&  
	\frac {1}{2}     \!  
	\left \{  
	\left[
	\left(  
	i\hbar
	\nabla
	- m{{\mbox{\bf v}}}({\mbox{\bf x}},t)
	\right)  
	{\hat \psi}^{\scriptscriptstyle\dagger}({\mbox{\bf x}})  
	\right]  
	{\hat \psi}({\mbox{\bf x}})
	-   \right.
	\\
	&&
	\hphantom{\frac {1}{2}     \!  
	\left \{  \right.
	}
	\left.
	{\hat \psi}^{\scriptscriptstyle\dagger}({\mbox{\bf x}})
	\left(  
	i\hbar \nabla + m{{\mbox{\bf v}}}({\mbox{\bf x}},t)
	\right)  
	{\hat \psi}({\mbox{\bf x}})  
	\right \}     \nonumber
	\end{eqnarray}
	by the relation:
	\begin{equation}
	\label{1.9}
	\langle\hat{\bf p}({\bf x})\rangle_t =
        \langle\hat m({\bf x})\rangle_t\cdot{\bf v}({\bf x},t)
	\end{equation}
	or equivalently
	\begin{equation}
	\label{1.10}
	\langle\hat{\bf p}^{(0)}({\bf x})\rangle_t =0
	\end{equation}
	The other classical state parameters are linked to the expectations
$\langle \hat e^{(0)}({\bf x}) \rangle_t$ 
and $\langle \hat m^{(0)}({\bf x})\rangle_t$
 (or $\langle\hat f^{(0)}({\bf x},{\bf p})\rangle_t$ 
in the kinetic case), as they enter into the
 structure of the most unbiased statistical operator, 
giving these assigned expectations. Such
 operator is characterized by the conditions:
 \begin{eqnarray}
 \label {1.11}
 \langle\hat e^{(0)}({\bf x} \rangle_t=\rm{Tr}(\hat e^{(0)}({\bf x})\hat \rho),
  \qquad
 \langle \hat m^{(0)}({\bf x}) \rangle_t=\rm{Tr}(\hat m^{(0)}({\bf x})\hat\rho)
 \end{eqnarray}
 \[
  0=\rm{Tr}( \hat {\bf p}^{(0)}({\bf x})\hat\rho)
 \]
$ S(\hat\rho)=-\rm{Tr}(\hat\rho log \hat\rho) $ being maximal.
The solution of this problem of conditioned maximality is a generalized Gibbs
 state
        \begin{equation}
        \label{1.12}
        {\hat w}[\beta(t),\mu (t),{{\mbox{\bf v}}}(t)]
	=  
        \frac{
        e^{
        -{
        \int_{\omega} d^3\! {\bf \scriptscriptstyle x} \,
	\beta({\bf \scriptscriptstyle x},t)  
        \left[
        {\hat e}^{(0)}({\bf \scriptscriptstyle x})
	-  
        \mu({\bf \scriptscriptstyle x},t){\hat m}^{(0)}
        ({\bf \scriptscriptstyle x})
	\right]  
        }}}
        {
	{\mbox{{\rm Tr}}} \, 
        e^{
        -{
	\int_{\omega} d^3\! {\bf \scriptscriptstyle x} \,  
	\beta({\bf \scriptscriptstyle x},t)  
	\left[  
	{\hat e}^{(0)}({\bf \scriptscriptstyle x})  
	-  
        \mu({\bf \scriptscriptstyle x},t){\hat m}^{(0)}
        ({\bf \scriptscriptstyle x})
	\right]  
        }}}\, ,
        \end{equation}
	where the fields $\beta({\bf x},t),\mu({\bf x},t)$ are determined
	by the equations:
	\[
        \langle {\hat e}^{(0)}({\bf x})\rangle_t=
        \rm{Tr}( {\hat e}^{(0)}({\bf x})
	{\hat w}[\beta(t),\mu(t),{\bf v}(t)])
	\] ,
	\[
        \langle{\hat m}^{(0)}({\bf x})\rangle_t=
        \rm{Tr}({\hat m}^{(0)}({\bf x})\hat w[\beta (t),\mu(t),{\bf v}(t)])
	\]
	in terms of the assigned expectations 
         $\langle {\hat e}^{(0)}({\bf x})\rangle_t$,
	${\hat m}^{(0)}({\bf x}) \rangle_t$ 
        and the given field ${\bf v}({\bf x},t)$ 
        (for simplicity the dependence of  ${\hat e}^{(0)}({\bf x})$
	on this field has not been made explicit).\\
In the kinetic case ${\hat m}^{(0)}\to {\hat f}^{(0)}$ and $\mu({\bf 
x},t)\to\mu({\bf {x}},{\bf {p}},t)$.	
The state function
	\[
	S(\beta(t),\mu(t),{\bf v}(t))
         =\rm{Tr}(\hat w[\beta(t),\mu(t),{\bf v}(t)]
         log \hat w[\beta(t),\mu(t),{\bf v}(t)])
         \]
	 is the thermodynamic entropy of the system.\
	 Let us take at the moment for granted that the dynamics 
         of the system is given by
	 the unitary evolution, generated by the Hamiltonian:
	 \begin{equation}
	 \label{1.13}
	 \hat H =\int_{\omega} d^{3} {\bf x}  \hat e({\bf x})
	 \end{equation}
         then the main point is to give at some time $T$ the initial 
	 statistical
	 operator $ \hat\rho_T$. Let us investigate what happens if one takes:
	 \begin{equation}
	 \label {1.14}
	 \hat\rho_T=\hat w[\beta(T),\mu(T),{\bf v}(T)]
	 \end{equation}
         then by a straightforward calculation (Robin, 1990),
	  the statistical operator of the system is:
	  \[
         \hat \rho_t=e^{-\frac{i}{\hbar}\hat H(t-T)}
          \hat \rho_T e^{\frac{i}{\hbar}\hat H(t-T)} \! ,
	  \]
          which can be written in the following form:
          \begin{equation}
	  \label{1.15}
          \hat \rho_t   =\frac{e^{-\langle \beta(t) 
          \cdot {\hat e}^{(0)} \rangle +
	  \langle [\beta(t) \mu(t)]\cdot {\hat m}^{(0)}\rangle+ \int_T^t d \tau
          {\hat S}_t[\beta(\tau),\mu(\tau),{\bf v}(\tau)]}}
          {\rm{Tr} e^{-\langle\beta(t)\cdot{\hat e}^{(0)}\rangle+
	  \langle[\beta(t)\mu(t)]\cdot{\hat m}^{(0)}\rangle+\int_T^t
        \!\!d\tau \hat S_t \left[ \beta(\tau),\mu(\tau),{\bf v}(\tau) \right] }}
          \end{equation}
The first two terms in the exponent are a more compact notation 
to represent the typical exponent
of a Gibbs state with parameters $\beta(t)$, $\mu(t)$, ${\bf v}(t)$ 
(the latter is implicit inside ${\hat e}^{(0)}$)
referring to time $t$, e.g.:
\[
\langle \beta(t)\cdot{\hat e}^{(0)}\rangle=\int_{\omega} d^3 {\bf x}
 \beta({\bf x},t)\hat e^{(0)}({\bf x},{\bf v}(t))
\]
The last term contains the history of the classical state parameters
 for $\tau\in[T,t]$:
 \begin{eqnarray}
 \label{1.16}
\lefteqn{ {\hat S}_t[\beta(\tau),\mu(\tau),{\bf v}(\tau)]=}     \nonumber\\
&& =\int_{\omega}d^3 {\bf x}
\left( \frac{\partial \beta({\bf x},\tau)}{\partial\tau}
\hat e({\bf x},\tau-t)-\nabla\beta({\bf x},\tau)\cdot
 \hat {\bf J}_e ({\bf x},\tau-t) \right) +{}  \nonumber  \\
&&{}
 +\int_{\partial\omega} d \sigma
{\bf n}\beta({\bf x},\tau) \hat{\bf J}_e({\bf x},\tau-t)+\cdots
 \end{eqnarray}
where $\hat{\bf J}_e$ is the energy current:
\[
\frac{i}{\hbar}[\hat H,\hat e({\bf x})]=-\rm{div} \hat{\bf J}_e({\bf x})
\]
the time dependence of the operators, e.g.: $\hat e({\bf x},t)$ means time
dependence in Heisenberg picture: $\hat e({\bf x},t)=e^{\frac{i}{\hbar}\hat H t}
\hat e({\bf x})e^{-\frac{i}{\hbar}\hat H t}$; at the r.h.s. of 
equation (\ref{1.16})
similar terms related to the other corresponding densities 
$ \hat m({\bf x})$, $\bf p({\bf x})$ have been omitted for brevity.\
In the framework of information thermodynamics expression (\ref{1.15}) 
would already be taken as a reliable
description of the system: 
no wonder at all about the different structure of $\hat\rho_T$ and $\hat\rho_t$
since at time $t$ information on the history interval $[T,t]$ 
is available, while it was not for $t<T$.
In our philosophy instead, $\hat\rho_t$ is an objective representation 
of the preparation of the system
until time $t$; expression  (\ref{1.15}) 
indicates that the history can
 be relevant, so the choice (\ref{1.14}) becomes highly critical: if the history
  of the system for $t<T$ is relevant, as one can expect looking at 
expression (\ref{1.15}),
   the choice (\ref{1.14}) (which is the most unbiased by the previous history)
    is wrong and also ${\hat\rho}_t$ given by (\ref{1.15}) is meaningless. 
A way out could be to shift
 $T\rightarrow -\infty$, thus eliminating the previous history:
  however the infinite system limit must be taken before, just the contrary
 of our attitude; furthermore the classical parameters for remote times are not
  a practically available input;
 in this way, associating to $\hat S_t[\cdots]$ in (\ref{1.15}) a
  factor $e^{(\tau-t)\epsilon}
 $ and taking $T=-\infty$ one obtains Zubarev's non-equilibrium statistical
 operator (Zubarev 1974).\\
 We propose another solution to the question whether 
(\ref{1.15}) makes sense: by the very
 definition of the classical parameters, the history term in 
(\ref{1.15}) is by
  construction irrelevant to calculate the expectations of the basic quantities
   $\hat e$ , $\hat m$ ,  $\hat f$ and  $\hat {\bf p} $ ; one can expect
   that its contributions to the expectations of the corresponding
    {\sl time derivatives}, e.g.:

   \[
   \dot{\hat e}=\frac{i}{\hbar}[\hat H,\hat e]=-\rm{div} \hat{\bf J}_e
    \]
    is small enough to allow a perturbative expansion of the exponential
   in (\ref{1.15}) with respect to the history term; then the classical
    parameters at time  $ \tau$
   contribute to correlation functions of the type flow-flow or flow-density,
    e.g.:
   $\langle \hat{\bf J}_e({\bf x}),\hat e({\bf y}, \tau -t)\rangle$.

The short-time behavior of such correlation functions is rapidly decaying,
when the time
separation of the two functions becomes of the order of a 
suitable decay time $\tau_c$.
Therefore if one considers $t>\tau_c$ only the part of the history 
referring to times $\tau> t-\tau_c$
does appreciably contribute to the dynamics of the relevant variables. 
Then to compute their dynamical behavior for times
$t$,  $t-T>\tau_c$, the initial condition
 (\ref{1.14}) is indeed the appropriate choice and
consequently $\hat\rho_t$ given by (\ref{1.15}) can be used to calculate 
the expectations
of the time derivatives, thus yielding closed integrodifferential evolution
equations for the classical variables
\[
z(t)\equiv( \beta(t),\mu(t), {\bf v}(t)), \qquad    t\geq T+\tau_c
\]
\begin{eqnarray}
\label{1.17}
\rm{Tr} \left( \frac{i}{\hbar}[\hat H,\hat e({\bf x})]\hat{\rho}_t \right)=
\frac{d}{dt}\rm{Tr} \left( \hat e({\bf x})
\hat w(\beta(t),\mu(t),{\bf v}(t)) \right) \nonumber \\
\rm{Tr} \left( \frac{i}{\hbar} [\hat H,\hat m({\bf x})]\hat{\rho}_t \right)=
\frac{d}{dt}\rm{Tr} \left( \hat m({\bf x})\hat w(\beta(t),
\mu(t),{\bf v}(t)) \right)  \\
\rm{Tr} \left( \frac{i}{\hbar}[\hat H,\hat{\bf p}({\bf x})]\hat{\rho}_t \right)=
\frac{i}{\hbar}\rm{Tr} \left(\hat {\bf p}({\bf x})
\hat w(\beta(t),\mu(t),{\bf v}(t)) \right)= \nonumber \\
=  \frac{d}{dt} \left( {\bf v}({\bf x},t)\rm{Tr}
( \hat m({\bf x}) \hat w(\beta(t),\mu(t),{\bf v}(t))) \right ) \nonumber
\end{eqnarray}
With respect to these equations, the values of the state variables 
$z(t)$ within the time
interval $[T , T+{\tau}_c]$ are prescribed parameters, related to the
 expectations of $\hat e({\bf x})$  ,
$\hat m({\bf x})$ and $ {\bf p}({\bf x})$: 
these expectations have the role of an input for the dynamics
at times $ t > T+{\tau}_c$. \
The time interval $[T , T+{\tau}_c]$ will be called 
{\sl {preparation time interval}} of the system; during such time interval one
might also assume that, due to the transition from
 $\hat {\psi}_{nc}$ to $\hat \psi$,
the flows indicated in equation (\ref{1.16}) do not vanish on the 
boundary $\partial\omega$, so that
a surface contribution can arise.
Neglecting mathematical problems about the existence of the solution of these
evolution equations, it seems at first that one has solved in 
principle the question of
a classical characterization of the finite isolated system: the parameters
 $z(t)$ establish the mathematical structure of the statistical 
operator that provides for times
 $ t > T+{\tau}_c$ the expectations of the relevant observables. However
 a serious flaw is evident at long times:\
 due to the pure point spectrum of $\hat H$  for a confined system, 
correlation functions have a quasiperiodical
 behavior, so they cannot decay indefinitely and as soon as memory is
recovered,
 choice (\ref{1.14}) is no longer tenable and then also 
$\hat{\rho}_t$ looses its meaning. Practically the difficulty can be
 avoided if one approximates expression (\ref{1.15}), replacing the integral
 $\int_T^t d\tau\cdots $ by $\int_{t-{\tau}_c}^td\tau \cdots$ and reproposing
 at time $t-{\tau}_c$ the initial condition
 \[
 \hat{\rho}_{t-{\tau}_c}\equiv \hat 
w[\beta(t-{\tau}_c),\mu(t-{\tau}_c),{\bf v}(t-{\tau}_c)]
 \]
 but this is to resort to an expedient.\
 However it is the basic assumption of unitary dynamics for the 
isolated system that leads to this difficulty and just this assumption 
becomes questionable
 inside the framework which sets isolated systems as basic elements of reality.
 \par
\section{An opening to irreversibility}
\setcounter{equation}{0}   
\par
 As it is clear from Section 2 the preparation of a finite system 
is described by
 a statistical operator, in our aim bearing the classical state 
parameters of the system;
 instead the quantum state vector $\psi \in {\cal H} $, 
bearing indexes related to measurements
 of observables is a very strong idealization which applies to highly controlled
 preparations typical of particle physics. If preparations are represented by
 statistical operators, i.e. elements of the set ${\cal K}$, the base of the
  positive cone in the space $T({\cal H}) $ of trace-class operators
 in ${\cal H}$, it becomes most natural to describe transformations of
  preparations by positive trace-preserving maps 
${\cal A}$ on $T({\cal H})$, these maps taking now the role
  that unitary transformations play in the theory based on 
microsystems. If such a map ${\cal A}$
  has an inverse one can show that ${\cal A}\cdot=\hat U\cdot\hat U^\dag$ with
  $\hat U$ unitary or antiunitary on ${\cal H}$ and in 
this way the two formulations become equivalent;
  but ${\cal A}$ need not have an inverse. 
Denoting by $\hat{\rho}_t$ a preparation
  performed until time $t$, let us consider for a system 
isolated during the time interval
  $[t_0 , t_1]$, the family of spontaneous repreparations 
$\rho_t$, $t\in[t_0,t_1]$, which
  arise due to the time evolution. One assumes that two preparations
  $\hat{\rho}_{t'}$,  $\hat {\rho}_{t''}$, 
$t_0\leq t'\leq t''\leq t_1$ are connected to each other
  by a map ${\cal A}(t''-t')$:
  \begin{equation}
  \label{2.1}
  \hat {\rho}_{t''}={\cal A}(t''-t')\hat{\rho}_{t'}
  \end{equation}
the family ${\cal A}(\tau) , \tau\ge0$ being a semigroup of positive,
trace-preserving maps. Actually taking the construction of section 2
 into account, since one restricts to the relevant
densities $\hat e({\bf x})$  , 
$ \hat {\bf p}({\bf x})$  , $ \hat m({\bf x})$  $ (\hat f({\bf x},{\bf p}))$,
looking at the time evolution in the Heisenberg picture, the properties of
${\cal A}'(\tau)$, which are mappings on ${\cal B}({\cal H})$
 into ${\cal B}({\cal H})$,
are important and one could assume only that ${\cal A}'(\tau)$ maps positive
densities into positive operators and conserves basic constants of motion
 like mass and energy:
${\cal A}'(\tau) \hat M=\hat M$, $ {\cal A}'(\tau) \hat H=\hat H$.\
Thus the basic feature of the description  characterized by {\sl { systems 
first and particles afterwards}} is irreversibility.\
In a sense we are now exploiting the arrow of time that is implicitly contained
in the operative approach to QM, based on preparations and measurements
 of prepared systems(Bohm, 1993). On the
other hand this almost trivial insertion of irreversibility into the formalism
 of
QM can easily be pushed back, as it is shown in Ludwig's approach
 to QM of microsystems: in his approach
a statistical operator represents an equivalence class of preparation
 procedures of a microsystem and
an {\sl {effect operator}} $\hat F,\quad   (0\leq\hat F\leq\hat1)$ represents an
 equivalence class
of registration procedures; any time shift of these procedures is allowed
 and still
one assumes that for any preparation procedure another one, shifted back
 in time, can be
found, equivalent to it. Then one arrives at a  unitary representation of
 time shifts
(Comi et al., 1975) and as a consequence, at unitary time evolution generated
by the Hamiltonian; then also the fundamental principle of conservation
 of energy is most directly settled.\
 The strategy we are proposing is to start always with 
local universal microphysics
 related to unitary representations of the fundamental symmetries, 
i.e. one has reversibility, energy conservation
 and some model leading to an energy density; 
e.g. in this preliminary discussion: expression
 (\ref{1.1}).\\
 Then, as we did in Section 2, one turns to the description of a system,
  characterized
 by a suitable choice of relevant variables.\
 Let us specialize our macroscopic system to the case of a dilute gas with
 short range interaction $V(|{\bf r}|)$; it is well known that for such a
 system a Boltzmann type of description is satisfactory: this description
  is characterized by a typical macroscopic variation time $\tau_1$ much larger
  than the duration of a collision: the mean free   path is much larger than
   the range of $V(|{\bf r}|)$. We shall see in Section 3
  that the introduction of the time scale ${\tau}_1$, i.e.: 
the detailed dynamics of the two-body interaction
  is replaced by collision, leads from the energy density (\ref{1.1}) to a 
semigroup 
${\cal A}'(\tau)$
  of maps, that display a stronger form of positivity, 
called {\sl {complete positivity}}
  which is reminiscent of the unitarity of the dynamics we started with; 
however this positivity will be
  {\sl relative} to the relevant variables.\
  We expect that the more general description based on equations 
(\ref{1.17})
  can be settled starting with a more fundamental expression than 
(\ref{1.1}).
  Here $V(|{\bf x}|)$ is a phenomenological input that 
could be derived from a Hamiltonian
  describing the structure of the molecule as a bound state of charged
  particles: i.e. the very presence of $V(|{\bf r}|)$ indicates that more
   fundamental fields should be considered.
  When dealing with the Hamiltonian dynamics of the charged fields one 
introduces a time scale
  typical of center of mass motion of the bound states, one expects that a
   semigroup $ {\cal A}'(\tau)$
  can be derived, whose generator displays an irreversible contribution together
  with the Hamiltonian contributions like (\ref{1.1}). 
In this way the difficulty we met in Section 2 with 
the long time behavior of the correlation
  function, should eventually disappear.
  \par
\section{Introduction of time scale and scattering map}
\setcounter{equation}{0}    
\par
  The relevant variables of the hydrodynamic or kinetic description 
have the following
  general structure, cf. Section 2:
  \[
  \sum_{hk}{\hat a}_ h^{\dag}A_{hk}(\xi) \hat a_k     ,    \sum_{k_1k_2h_1h_2}
 {\hat a}_{h_1}^{\dag}{\hat a}_{h_2}^{\dag}
A_{h_1h_2k_2k_1}({\bf x})\hat a_{k_2}\hat a_{k_1}
 \ ;
  \]
thus, in Heisenberg picture, we have to study expressions of the form:
\begin{equation}
\label{3.1}
\sum_{hk}e^{\frac{i}{\hbar}\hat H t}\hat a_h^{\dag}
\hat a_k e^{-\frac{i}{\hbar}\hat H t}A_{hk}(\xi) \quad
\sum_{k_1k_2h_1h_2}e^{\frac{i}{\hbar}\hat H t}
\hat a_{h_1}^{\dag}\hat a_{h_2}^{\dag}
\hat a_{k_2}\hat a_{k_1}e^{-\frac{i}{\hbar}\hat H t} 
A_{h_1h_2k_2k_1}({\bf x})\      ;
\end{equation}
where restriction to slow variables means that in the 
sums (\ref{3.1}) only terms are considered
such that:
\begin{equation}
\label{3.2}
\frac{1}{\hbar}|E_h-E_k|<\frac{1}{\tau_1} \qquad
   \frac{1}{\hbar}|E_{h_1}+E_{h_2}-E_{k_1}-E_{k_2}|< \frac{1}{\tau_1}
\end{equation}
where $\tau_1$ is the typical variation time of the 
Boltzmann description: $\tau_1
\sim 10^{-13} \textrm{sec}$, the time interval between two collisions.\
The Hamiltonian $\hat H$ , given by (\ref{1.13}) generates an isomorphism
${\cal U}_H'(t)$ of ${\cal B}({\cal H})$:
\begin{equation}
\label{3.3}
{\cal U}_H'(t)\cdot = 
e^{\frac{i}{\hbar}\hat Ht}\cdot e^{-\frac{i}{\hbar}\hat Ht}=
\int_{-i\infty+\epsilon}^{i\infty+
\epsilon}dz\frac{e^{zt}}{2\pi i}\frac{1}{z-{\cal H}'}       ,
\end{equation}
where ${\cal H}'\cdot =\frac{i}{\hbar}[\hat H, \cdot]\! $ .\
We shall introduce a formalism reminiscent of usual 
scattering theory shifting the space
${\cal H}$ to ${\cal B}({\cal H})$ and operators in 
${\cal H}$ to maps in ${\cal B}({\cal H})$; for
brevity only the main steps of the treatment are indicated:
\begin{equation}
\label{3.4}
        {{
	{1\over{ z - {\cal H}^{'}}}  
	}}  
	=  
	{{  
	{1\over{ z - {\cal H}^{'}_0}}  
	}} +{{  
	{1\over{ z - {\cal H}^{'}_0}}  
	}}  
	{\cal T}(z){{  
	{1\over{ z - {\cal H}^{'}_0}}  
	}}
 \end{equation}
 \begin{equation}
 \label{3.5}
                {\cal T}(z)
	\equiv  
	{\cal V}^{'} + {\cal V}^{'}{{  
	{1\over{ z - {\cal H}^{'}}}  
        }}{\cal V}^{'}      ,
 \end{equation}
where ${\cal H}_0'=\frac{i}{\hbar}[\hat H_0 , \cdot]$   , 
$ {\cal V}'=\frac{i}{\hbar}[\hat H-\hat H_0,\cdot]$ ,
$\hat H_0=\sum_fE_f \hat a_f^{\dag}\hat a_f$.\
The operators
\begin{equation}
\label{3.6}
\hat a_{h_1}^{\dag}\hat a_{h_2}^{\dag}
\cdots\hat a_{h_r}^{\dag}\hat a_{k_s}\cdots
\hat a_{k_2}\hat a_{k_1}
\end{equation}
are {\sl eigenstates} of ${\cal H}_0'$ with eigenvalues 
$\frac{i}{\hbar}(E_{h_1}+E_{h_2}
\cdots +E_{h_r}-E_{k_1}-E_{k_2}-\cdots-E_{k_s})$.\
By the basic algebraic property:
\begin{equation}
\label{3.7}
{\cal U}'(t) \hat a_h^{\dag}\hat a_k = \left({\cal U}'(t)\hat a_h\right)^{\dag}
\left({\cal U}'(t)\hat a_k\right)        ,
\end{equation}
it is clear that the main formal tool to treat expressions 
\ref{3.1} is the representation of the operator
${\cal T}(z) \hat a_k$ , in terms of the basis \ref{3.6}.\
By conservation of total mass one has the general structure:
\begin{eqnarray}
\label{3.8}
{\cal T}(z)\hat a_k=\sum_fA_{kf}(z,\hat n)\hat a_f+
\sum_{lf_2f_1}\hat a_l^{\dag}A_{lkf_2f_1}
(z,\hat n)\hat a_{f_2}\hat a_{f_1}+{}\nonumber \\
{}+\sum_{l_1l_2f_3f_2f_1}\hat a_{l_1}^{\dag}
\hat a_{l_2}^{\dag}A_{l_1l_2kf_3f_2f_1}(z,\hat n)
\hat a_{f_3}\hat a_{f_2}\hat a_{f_1}+\cdots
\end{eqnarray}
where the coefficients $A_{lkf_2f_1}(z,\hat n)$, 
$A_{l_1l_2kf_3f_2f_1}(z,\hat n)$ are
operator functions of the set of number operators 
$\hat n_h=\hat a_h^{\dag}\hat a_h$, i.e.:
they are diagonal with respect to the basis in Fock space, 
generated by the creation operators.
A very natural approximation in usual kinetic theory is the 
evolution by two-particle collisions. In our field description
the corresponding approximation seems to be the following: 
evolution involving {\sl only one} other field mode;
i.e. one would break up expansion (\ref{3.8}) after the first two 
terms; 
however all the {\sl spectator
modes} are also relevant through the $\hat n$ dependence of the 
coefficients and provide the Pauli principle
corrections. Then also the third term in (\ref{3.8}) 
contributes to these Pauli principle corrections inside
an expression of the form $\hat a_h^{\dag}{\cal T}(z)\hat a_k$ 
when some index $f_1,f_2,f_3$ is equal to $h$.
Let us indicate briefly the structure of the coefficient 
$A_{lkf_1f_2}(z,\hat n)$
in (\ref{3.8}); it is given essentially by the matrix elements 
of a two-particle scattering operator, bearing
Pauli-principle corrections, defined as follows:      
\begin{equation}
\label{3.9}
\hat T^{(2)}(z)=\hat V^{(2)}+ 
\hat V^{(2)} \frac{1}{z-\hat H_L^{(2)}}\hat V_L^{(2)}\quad ,
\quad \hat H_L^{(2)}=\hat H_0^{(2)}+\hat V_L^{(2)}
\end{equation}
the operators labeled by the index (2) are
defined in the Hilbert space ${\cal H}^{(2)}$ of two identical particles
by matrix elements in the two particle  
(symmetric or antisymmetric)  basis $|l_1l_2>$:
\begin{eqnarray}
\label{3.10}
<l_2l_1|\hat H_o^{(2)}|f_1f_2>=(E_{f_1}+E_{f_2})
\frac{1}{2!}(\delta_{l_2f_2}\delta_{l_1f_1}
\pm\delta_{l_2f_1}\delta_{l_1f_2}) \nonumber \\
<l_2l_1|\hat V^{(2)}|f_1f_2>=V_{l_1l_2f_2f_1}& & , \nonumber \\
 <l_2l_1|\hat V_L^{(2)}|f_1f_2>=(1\pm\hat n_{l_1}
\pm\hat n_{l_2})V_{l_1l_2f_2f_1} & &
\end{eqnarray}
These {\sl two-particle} quantum mechanical elements, 
are produced in a natural way by the quantum field structure and are constructed
with the coefficients $E_f$ ,   $V_{l_1l_2f_2f_1}$ , arising in the Hamiltonian
(\ref{1.13}) , written in terms of $ \hat a_l^{\dag} , \hat a_f $:
\begin{equation}
\label{3.11}
\hat H=\sum_f E_f \hat a_f^{\dag}\hat a_f^{\dag}+ \frac{1}{2}\sum_{l_1l_2f_1f_2}
\hat a_{l_1}^{\dag}\hat a_{l_2}^{\dag}V_{l_1l_2f_2f_1}\hat a_{f_2}\hat a_{f_1}
\end{equation}
The factor $(1\pm\hat n_{l_1}\pm\hat n_{l_2})$ in the last of 
equations ( \ref{3.10})
represents the Pauli-principle correction, it is an operator valued 
expression, but this makes no problem for the definition  
(\ref{3.9}) of ${\hat T}^{(2)}(z)$ since
for all $l_1 , l_2$ they commute.
We assume for simplicity that no bound states between the 
molecules can be formed, this means that in the thermodynamic
limit the coefficients in (\ref{3.8}) have no singularities on the 
imaginary axis.
Then the time scale is introduced by the following modifications. 
First the expression ${\cal T}(z)\hat a_k$ is replaced
by  ${\cal T}(z+\eta)\hat a_k$ , 
with $\eta\approx\frac{\hbar}{\tau_0} $  , $\tau_0$
being of the order of the collision time. 
Final results for expectations of the relevant variables, 
having a typical variation
time $\tau_1\gg\tau_0=\frac{\hbar}{\eta}$ are practically independent on $\eta$:
actually only these $\eta$ independent results are 
significant in our essentially incomplete description
of the finite system; $\eta$ dependence would mean dependence 
on the distribution of the huge set of poles that 
${\cal T}(z)$ has on the imaginary axis, 
which in turn is related to the confinement of the
system, only roughly represented by the boundary
conditions we assumed in Section 2; 
so $\eta$ dependence is more an artifact of the idealized
confinement than a physical feature.\
A second change concerns the external variables $E_h , E_k$ ; we set for them
$E_h=E_{hk}+\frac{1}{2}\xi_{hk}$, 
$E_k=E_{hk}-\frac{1}{2}\xi_{hk}$  and the replace 
$  \xi_{hk}\rightarrow \xi_{hk}-2i\epsilon$,
with $\eta>\epsilon>>\frac{\hbar}{\tau_1}$, 
thus implying some smoothness property of the
dependence on the variable $\xi_{hk}$, 
as the existence of an analytic continuation into the
lower half-plane.
Now expressions (\ref{3.7}) are calculated taking into account 
equations (\ref{3.3}), (\ref{3.4}) and the representation
(\ref{3.8}) where $z\rightarrow z+\eta$. 
Then one can separate in the calculation of (\ref{3.3}) the
contribution of the singularities of 
$\frac{1}{z-{\cal H}_0'}$ from the 
singularities at points $z$, with $ \rm{Re}\,z\leq -\eta$ and neglect
the last ones: 
their contribution is negligible 
for $t>> \tau_0$ if we consider only relevant variables. 
Finally one arrives with some  calculations to the following very perspicuous
 representation of expression (\ref{3.7}):
 \begin{displaymath}
 {\cal U}'(t) \hat a_h^{\dag}\hat a_k=
\hat a_h^{\dag} \hat a_k+t {\cal L}'(\hat a_h^{\dag}\hat a_k) \quad
 t>>\tau_o ,  \quad  t\sim \tau_1
 \end{displaymath}
 where
 \begin{equation}
 \label{3.12}
 {\cal L}'\hat a_h^{\dag}\hat a_k=\frac{i}{\hbar}\left[\hat H_{\rm{eff}}
  , \hat a_h^{\dag}\hat a_k \right]-
 \frac{1}{\hbar} \left( \left[\hat \Gamma , \hat a_h^{\dag}\right]\hat a_k-
  \frac{1}{\hbar} \hat a_h^{\dag} \left[ \hat\Gamma , \hat a_k \right] \right) +
 \frac{1}{\hbar}\sum_{\lambda}\hat R_{h\lambda}^{\dag}\hat R_{k\lambda}   ;
 \end{equation}
 In the first term at the r.h.s. of equation (\ref{3.12}) 
an {\sl effective} Hamiltonian appears, given
 by equation (\ref{3.11}) with $V_{l_1l_2f_2f_1}$ 
replaced by $ V_{l_1l_2f_2f_1}^{\rm{eff}}$,
 \begin{displaymath}
 V_{l_1l_2f_2f_1}^{\rm{eff}}= {}  \\
 {} = <l_2l_1|\frac{1}{2}\left(\hat T^{(2)}(E_{f_1}+E_{f_2}+i\eta\hbar)+
 \hat T^{(2)}(E_{l_1}+E_{l_2}+i\eta\hbar)^{\dag}\right)|f_2f_1>
 \end{displaymath}
 Thus the interaction potential is replaced by 
the self-adjoint part of the scattering operator;
 the remaining part of the scattering operator is not zero if one 
goes beyond the Born approximation
 and yields the second term at r.h.s. of equation (\ref{3.12}), 
which is no longer of the form
 $[\cdot,\hat a_h^{\dag}\hat a_k]$,
 \[
 \hat \Gamma = \frac{1}{2}\sum_{f_1f_2l_1l_2}\hat a_{l_1}^{\dag}
  \hat a_{l_2}^{\dag}
 <l_2l_1| \frac{i}{2}\left(\hat T^{(2)}(E_{f_1}+E_{f_2}+i\eta\hbar)
  -\hat T^{(2)}(E_{l_1}+E_{l_2}+i\eta\hbar)^{\dag}\right)|f_ 2f_1>
 \hat a_{f_2}\hat a_{f_1}
 \]
 The operators $\hat R_{k\lambda}$ are given by:
 \begin{equation}
 \label{3.13}
 \hat R_{k\lambda}=-i\sqrt{ 2\epsilon(1\pm\hat n_{\lambda}\pm\hat n_k)}
 \sum_{f_1f_2}\frac{<k\lambda|\hat T^{(2)}
(E_{f_1}+E_{f_2}+i\hbar(\eta- \epsilon))}
 {E_k+E_{\lambda}-E_{f_1}-E_{f_2})-i\hbar\epsilon} \hat a_{f_2}\hat a_{f_1}   ,
 \end{equation}
 the factor $\sqrt{2\epsilon(1\pm\hat n_{\lambda}\pm\hat n_k)}$
  arising by the approximation:
  \[
  2\epsilon(1\pm\hat n_{\lambda}\pm(\hat n_k\pm\hat n_k))\approx\sqrt{2\epsilon
   (1\pm\hat n_{\lambda}\pm\hat n_h)}
\sqrt{2\epsilon(1\pm\hat n_{\lambda}\pm\hat n_k)}
   \]
   which holds in the case of not too large Pauli principle corrections.
   Because of mass conservation one has within our approximations:
   \begin{equation}
   \label{3.14}
   \hat \Gamma^{(2)}\approx\frac{1}{4}
\sum_{h\lambda}\hat R_{h\lambda}^{\dag}\hat R_{k\lambda}
   \end{equation}
   or in order to have exactly ${\cal L}'\hat M=0$, one can set in 
equation (\ref{3.12}):
   \[
   \hat \Gamma^{(2)}=
\frac{1}{4}\sum_{h\lambda}\hat R_{h\lambda}^{\dag}\hat R_{h\lambda}\   .
   \]
    Let us notice that the general structure of 
${\cal L}'$, arising from the factorized form shown by (\ref{3.7}),
    indicates a form of complete positivity {\sl relative} to the
     operators $\hat a_h^{\dag}\hat a_k$; one can easily see that:
     \[
     0\leq\sum_{hk}<\psi_h|([1+t\cdot{\cal L}')
\hat a_h^{\dag}\hat a_k|\psi_k>  \quad \forall 
\quad \{ {\psi_h}\}     \subset {\cal H}
     \]
     to first order in $\tau$.\
     Assuming (\ref{1.1}) as {\sl fundamental} energy density, 
introducing confinement and a time scale much larger than
     $\tau_0$, we arrived by a systematic procedure to the 
generator ${\cal L}'$: 
the effective Hamiltonian is now associated with a non 
Hamiltonian contribution. By a similar treatment
     for the system consisting of a particle interacting 
with a medium, one can obtain ${\cal L}'$
     for the particle variables (Lanz, Vacchini, 1997), then considering
     the time evolution of the statistical 
operator for the sole particle, the typical
     quantum master equation describing Brownian motion is obtained.\\
      In our present case we can expect that, on
     the time scale ruled by ${\cal L}'$, 
no memory of the classical state variables introduced 
in Section 2 is relevant, so
     that one has a closed set of evolution equations:
     \begin{equation}
     \label{3.15}
     \rm{Tr}\left( ({\cal L}'\hat f({\bf x}))
\hat w(\beta(t),\mu(t),{\bf v}(t) \right)=
     \frac{d}{dt}\rm{Tr}(\hat f({\bf x})\hat w(\beta(t),\mu(t){\bf v}(t))
     \end{equation}
     where $\hat f({\bf x})$ are the relevant fields  $\hat e({\bf x})$ ,
      ${\bf p}({\bf x})$ , $\hat m({\bf x})$.
     Looking at ${\cal L}'(\hat a_h^{\dag}\hat a_h)$ one can see that
     $\sum\hat R_{h\lambda}^{\dag}\hat R_{h\lambda}$ has the typical structure
      of the gain contribution by a collision ending up in the
     two-particle state $h\lambda$ ,which is present in the Boltzmann collision
      term, while $-\frac{1}{\hbar}([\Gamma,\hat a_h^{\dag}]\hat a_h
       -\hat a_h^{\dag}[\hat \Gamma, \hat a_h])$
     yields the loss term by a collision involving a particle in the state
      $h$.\
     Therefore one can expect that the 
description based on equations (\ref{3.15}) is an improvement of 
the usual Boltzmann equation,
     because no factorization 
hypothesis of two-particle distribution function must be used
     and $\hat H^{\rm{eff}}$ is not purely kinetic. 
We hope that the procedure used
     to obtain  ${\cal L}'$, 
which is based on smoothness properties of the scattering map ${\cal T}(z)$
     can be extended to the case of singularities of ${\cal T}(z)$ related to
     bound states: so one could also make 
that the energy density (\ref{1.1}) from which we started, fits in a 
suitable ${\cal L}'$, derived by a more fundamental model.
\vskip 100pt
{\parindent = 0 pt
{\LARGE References}
\vskip 15pt
{Bohm, A.}
(1993).
Gamow vectors and the arrow of time,
in
{\it Symposium On the Foundations of Modern Physics}, (1993)
Busch, P., Lahti, P.~J., and Mittelstaedt, P., eds., World
Scientific, Singapore, p.77-97

{Comi, M.,  Lanz, L.,  Lugiato, L.A., and Ramella, G.,}
(1975) 
{\it Journ. Math. Phys.}
{\bf 16},
{910}

{Davies, E.~B.}
(1976).
{\it Quantum Theory of Open
Systems}, {Academic Press}, {London}.

{Holevo, A.~S.}
(1982).
{\it Probabilistic and Statistical Aspects of
Quantum Theory}, {North Holland}, {Amsterdam}{}.

{Kraus, K.}
(1983).
States, Effects and Operations, in {\it Lecture Notes in
Physics}, Volume 190, {Springer}, {Berlin}.

{Lanz, L., Melsheimer,  O. and Wacker, E.}
(1985)
{\it Physica},
{\bf 131A},
{520}

{Lanz, L., and Vacchini, B.}
(1997).
{\it International Journal of Theoretical Physics},
{\bf 36},
{67}.

{Ludwig, G.}
(1983).
{\it Foundations of Quantum
Mechanics}, {Springer}, {Berlin}.

{Robin, W.~A.}
(1990).
{\it Journal of Physics A},
{\bf 23},  
{2065}.
  
{Zubarev, D.~N.}
(1974).
{\it Non-equilibrium statistical thermodynamics},
Consultant Bureau, New York.
}
\end{document}